\documentclass{iopart}
\usepackage[dvips]{graphicx}
\usepackage{amssymb}

\newcommand{\be}{\begin{equation}}
\newcommand{\ee}{\end{equation}}
\newcommand{\ba}{\begin{eqnarray}}
\newcommand{\ea}{\end{eqnarray}}

\newcommand{\bml}{\begin{mathletters}}
\newcommand{\eml}{\end{mathletters}}
\def\ltsima{$\; \buildrel < \over \sim \;$}
\def\simlt{\lower.5ex\hbox{\ltsima}}
\def\gtsima{$\; \buildrel > \over \sim \;$}
\def\simgt{\lower.5ex\hbox{\gtsima}}

\begin{document}

\title[LISA Response Function and Parameter Estimation]
{LISA Response Function and Parameter Estimation}

\author{Alberto Vecchio\footnote{av@star.sr.bham.ac.uk} and Elizabeth D.~L.~Wickham  
\footnote{edlw@star.sr.bham.ac.uk}
}

\address{School of Physics and Astronomy, The University of Birmingham, Edgbaston Birmingham B15 2TT, UK}

\begin{abstract}
We investigate the response function of LISA and consider the adequacy of its commonly used approximation in the high-frequency range of the observational band. We concentrate on monochromatic binary systems, such as white dwarf binaries. We find that above a few mHz the approximation starts becoming increasingly inaccurate. The transfer function introduces additional amplitude and phase modulations in the measured signal that influence parameter estimation and, if not properly accounted for, lead to losses of signal-to-noise ratio.
\end{abstract}

\section*{PACS numbers: 04.30.-w, 04.80.Nn, 97.20.Rp, 97.80.-d}




\section{Introduction}
The Laser Interferometer Space Antenna (LISA) is a collaborative ESA/NASA enterprise
to fly a space borne laser interferometer of arm-length 5 million km for 
the observation of gravitational waves (GW's) in the frequency window $10^{-4}$ Hz - 
0.1 Hz~\cite{lisa_ppa}. LISA is likely to observe
the largest sample of galactic binary systems of stellar-mass compact objects,
strongly dominated by double white dwarfs (WD's), but including also neutron stars
and possibly solar-mass black holes~\cite{NYP01}.
Due to the very large sample of these (electro-magnetically very faint) objects,
it is therefore interesting to investigate new astrophysical information that LISA
will be able to provide, and therefore estimate the errors associated with parameter 
measurements. All the investigations that have been carried out so far regarding
short-period stellar mass binary systems have
implicitly assumed LISA to work in the so-called long
wavelength approximation (which is equivalent to assuming that
the instantaneous GW emission frequency is $f\ll 10^{-2}$ Hz), in other words that 
the LISA transfer function is
constant~\cite{Peterseimetal96,Peterseimetal97,Cutler98,CV98,TN02,Seto02}.
However the vast majority of the detectable systems -- current estimates suggest
that LISA will be able to resolve $\sim 10^4$ sources -- will be in the
frequency band 1 mHz - 10 mHz~\cite{NYP01}, where the frequency and location
dependent transfer function might introduce significant effects; such effect has already been noted for the case of observations of black hole binary inspirals~\cite{Seto03}.

Here we address the effect of LISA's finite arm-length for parameter estimation
in observations of monochromatic signals, and compare the results with those
from previous analyses.

\section{Parameter estimation}

The signal at the LISA Michelson detector output reads~\cite{CL01}
\begin{eqnarray}
h(t) & = & D^{ab}(t)h_{ab}(t)\,,
\label{h}
\end{eqnarray}
where $h_{ab}(t)$ describes the metric perturbation induced by GW's,
and
\begin{eqnarray}
D^{ab}(t)&=&\frac{1}{2}\left[l^{a}_{j}(t)l^{b}_{j}(t)T_{j}(t)- 
l^{a}_{k}(t)l^{b}_{k}(t)T_{k}(t)\right]
\label{D}
\end{eqnarray}
is the detector response tensor.
In the previous expression ${l}^{a}_{j}$ 
is the unit vector along the $j$-th arms of the interferometer
($j = 1,2,3$ for the three arms of LISA here we simply
consider one detector with $j$=1 and $k$=2)
and 
\begin{eqnarray}
T_{j}(t) & = & \frac{1}{2}\,\mathrm{sinc}
\left[\frac{f}{2f_{\ast}}(1+{l}^a_{j} N_a)\right]
\,\exp\left\{-i\left[\frac{f}{2f_{\ast}}(3-{l}^a_{j}{N}_a)\right]\right\} 
\nonumber \\
 & & +\frac{1}{2}\,\mathrm{sinc}\left[\frac{f}{2f_{\ast}}(1-{l}^a_{j}{N}_a)\right]
\,\exp\left\{-i\left[\frac{f}{2f_{\ast}}(1-{l}^a_{j} {N}_a)\right]\right\}
\label{T}
\end{eqnarray}  
is the LISA transfer function which tends to 1 in the low frequency limit 
defined by $f/f_{\ast}\ll 1$;  $f_{\ast} = c/(2\pi L) \simeq 9.5$ mHz 
is the LISA transfer frequency ($L = 5\times 10^6$ km) and the unit vector $N^a$ identifies
the source location in the sky. We note the time dependence of $T_{j}(t)$ due 
to the change in orientation of LISA during the year long observation 
time~\cite{lisa_ppa}.

Setting $T = 1$ in Eq.(\ref{D}) and then substituting into Eq.(\ref{h}) 
returns the familiar expression of the signal
at the detector output in the long wavelength approximation~\cite{Cutler98},
\begin{eqnarray}
h_{\rm l}(t) = F^{+}(t)\,h^{+}(t) + F^{\times}(t)\,h^{\times}(t),
\label{hl}
\end{eqnarray}
where $F^{+}$ and $F^{\times}$ are the so-called antenna beam patterns and 
$h^{+}(t)$ and $h^{\times}(t)$ are the two independent polarisations. 
We henceforth refer to these two representations as $h$ for the exact response 
function and $h_{l}$ for the response function in the long wavelength 
approximation.

To establish at what frequency the two representations $h$ and
$h_{\rm l}$ become significantly different, we compute the overlap
\begin{eqnarray}
O =\frac{(h|h_{\rm l})}{\sqrt{(h|h)(h_{\rm l}|h_{\rm l})}}\,,
\label{O}
\end{eqnarray} 
where $(u|v)$ represents the usual inner product between two functions $u$ and 
$v$~\cite{Cutler98}.
Notice that in Eq~(\ref{O}) we do not maximise over any of the signal parameters.
We have calculated $O$ over the frequency range $10^{-4}$ Hz -- $10^{-2}$ Hz for randomly selected sky locations and orientations of the orbital plane, which is identified by the unit vector $L^{a}$ along the direction of the orbital angular momentum; we have found that $O\approx 0.9$ at $\approx 5$ mHz (the peak of the LISA sensitivity is at 3 mHz) and the overlap drops rapidly below $\approx 0.5$ for $f\sim 10^{-2}$ Hz, which suggests that at frequencies higher than a few mHz the low frequency approximation may no longer be appropriate; this has important repercussions for signal detection through matched filters, and we are currently investigating this issue in more detail.

To investigate the statistical errors associated with parameter estimation we compute the Fisher information matrix $\Gamma^{jk}$, and its inverse $(\Gamma^{jk})^{-1} = \langle\Delta\lambda^{j}\Delta\lambda^{k}\rangle $, which is related to the variance-covariance matrix through a straightforward normalisation of the off-diagonal elements~\cite{Cutler98}. We do this for both $h$ and $h_{\rm l}$. $\lambda^j$
represents the vector of the unknown parameters, in our case the 
signal amplitude
$A$, the constant frequency $f_0$, the arbitrary initial phase $\phi_0$ and
the four angles that define the source geometry with respect to the detector (in our
representation we use as independent parameters
$\cos\theta_{N}, \phi_{N}, cos\theta_{L}, \phi_{L}$).
We also evaluate the LISA angular resolution, defined as $
\Delta \Omega  = 2 \pi(\langle\Delta\cos\theta_N^2\rangle\langle\Delta\phi_N^2\rangle -
\langle\Delta\cos\theta_N\,\Delta\phi_N \rangle^2)^{1/2}$.

\begin{figure}[htbp]
  \vspace{3pt}
  \begin{center}
    \includegraphics[height=3.0in]{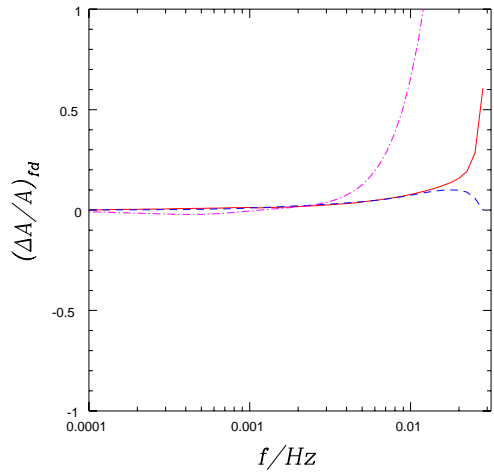}    
  \end{center}
  \caption{
   Comparison of the measurement errors. 
   The plot shows the fractional difference of 
$\Delta A/A$, Eq.(\protect{\ref{X}}), as a function of frequency 
   between the low-frequency approximation and the exact expression of the signal measured at the detector output for three randomly selected positions and orientations of a source in the sky: 
solid line,2.2,4.6,2.5,4.6;
dashed line,0.90,1.2,1.7,1.1;
dot-dashed line,2.8,4.0,0.89,3.9; angles are listed corresponding to 
$\theta_{N},\phi_{N},\theta_{L},
\phi_{L}$.}
  \label{errora}
\end{figure}

\begin{figure}[htbp]
  \vspace{3pt}
  \begin{center}
    \includegraphics[height=3.0in]{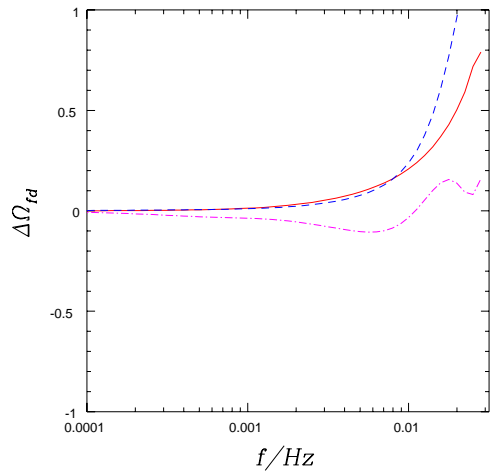}    
  \end{center}
  \caption{
   Comparison of the measurement errors. 
   The plot shows the fractional difference of $\Delta \Omega$, for the same
positions and orientations in the sky, c.f. Figure 1.}
  \label{erroro}
\end{figure}

We investigate the differences between the expected errors computed using the low frequency approximation and the exact expression by computing their fractional difference
\begin{eqnarray}
X_{\rm fd} = (X_{\rm l}-X)/X\,,
\label{X}
\end{eqnarray}
where $X$ stands for either the mean-square error or the angular resolution and the subscript $l$ indicates that the long wavelength approximation has been used. 
We have calculated $X_{\rm fd}$ over the LISA observational window for randomly selected positions and orientations in the sky. Here we show results for $\Delta A/A$ and $\Delta \Omega$. 
The results are reported in Figures~\ref{errora} and~\ref{erroro} and 
clearly show that the transfer function introduces 
characteristic modulations 
in phase and amplitude, c.f. Eq.(3)
 that help in resolving the parameters more accurately. 

Our preliminary results
indicate that LISA's angular resolution differs by $\approx 5\%$ and $\approx 30\%$
at 5 mHz and 10 mHz, respectively, with respect to the long-wavelength approximation.
Analogously, $\Delta A/A$ differs by a few percent and $\approx 20\%$, respectively. It is clear that the results depend strongly on the signal frequency and the source geometry, and $h_{\rm l}$ becomes indeed an increasingly poorer representation of the actual detector output for $f \simgt $ 5 mHz.


\section{Conclusions}
We have shown that in the frequency range 1 mHz - 10 mHz, where most of
the WD binary systems will be detected, the effect of the finite length of
LISA arms can be significant and needs to be properly taken into account in LISA
data analysis. Our analysis is limited in two main respects: (i) we have explored 
only a
limited portion of the whole signal parameter space, and (ii) we have not
allowed for intrinsic frequency drifts of the signal during the observation time.
The analysis to address the two former issues is currently in progress~\cite{VW}.

\end{document}